\documentclass[reprint, twocolumn, pre]{revtex4-1}

\pdfoutput=1

\usepackage{graphicx}

\begin{document}

\title{Comment on ``Capturing correlations in chaotic diffusion by
  approximation methods''} 
\author{Thomas Gilbert}
\email{thomas.gilbert@ulb.ac.be}
\affiliation{Center for Nonlinear Phenomena and Complex Systems,
  Universit\'e Libre  de Bruxelles, C.~P.~231, Campus Plaine, B-1050
  Brussels, Belgium}
\author{David P.~Sanders}
\email{dpsanders@ciencias.unam.mx}
\affiliation{Departamento de F\'isica, Facultad de Ciencias, Universidad
Nacional Aut\'onoma de M\'exico,  Ciudad Universitaria,
04510 M\'exico D.F.,
 Mexico}

\maketitle

In their paper \cite{KK}, Knight and Klages argue that the application to
their one-dimensional map model of the approximation scheme for the
characterization of transport coefficients of diffusion processes developed
by the present authors in Refs.~\cite{GS} is hampered by the presence of
states with zero velocity. This observation would seem like a serious
limitation of our formalism, but, as we illustrate below, it is in fact
inaccurate,  and is due rather to a misinterpretation of the Machta-Zwanzig
approximation \cite{MZ}, based on a confusion between two distinct timescales.

Indeed, consider a simple symmetric random walk on a one-dimensional
lattice with unit spacing $\ell = 1$ between neighboring sites, in which a
walker hops at every unit time step left or right with probability $p$,
$0<p<1/2$, and stays put with probability $q = 1 - 2p$. It is a trivial
calculation to show that the mean squared displacement per unit step of
such a walker is $2p$, so that the diffusion coefficient of this process is
simply $D = p$.

The same result is equally obtainable by application of the Machta-Zwanzig
formula \cite{MZ}. This dimensional prediction, according to which (in one
spatial dimension) the diffusion coefficient is one half of the square of
the length $\ell$ of the walker's displacements divided by their timescale
$\tau$, must indeed be exact for this simple model, since walkers have no
memory of their  history. Accordingly, the diffusion coefficient is the
lattice spacing squared (which we took to be unity), divided by twice the
timescale of jumps. That timescale is the average time spent by a walker at
any given site before it moves on to a neighboring one. In this case, it is
easily computed to be $\tau = 1 / (2p) > 1$.  This recovers the
previously-stated result. 

In other words, the diffusion coefficient of the symmetric random walk
described above is equal to that of another, yet simpler, process, in which
a walker jumps left or right with probability $1/2$ at rate $2p$ (whether
in continuous or discrete time). Ignoring the underlying dynamics, these
quantities are easily accessible to direct measurements, which is why this
prediction is so useful.

The flaw in the attempt of Knight and Klages \cite{KK} to transpose our
formalism \cite{GS} to their model is to have confused the timescale of
displacements with the unit timescale of the underlying
process. Zero-velocity states are in fact irrelevant, since they are
accounted for by plugging in the relevant timescale, that of
displacements. Our formalism is but a simple extension of the above
considerations, which allows for the systematic inclusion of memory effects
in such processes in a self-contained way. It is in fact very general and
is easily applied by direct measurement of the appropriate quantities.


\begin{thebibliography}{9}
\bibitem{KK} G. Knight and R. Klages, Phys. Rev. E \textbf{84}, 041135
  (2011).
\bibitem{GS} T. Gilbert and D. P. Sanders, Phys. Rev. E \textbf{80}, 041121
  (2009); T. Gilbert and D. P. Sanders, J. Phys. A:
  Math. Theor. \textbf{43}, 035001 (2010); T. Gilbert, H. C. Nguyen, and
  D. P. Sanders, J. Phys. A: Math. Theor. \textbf{44}, 065001 (2011).
\bibitem{MZ} J. Machta and R. Zwanzig, Phys. Rev. Lett. \textbf{50}, 1959
  (1983). 
\end{thebibliography}
\end{document}